\documentclass[journal=jpclcd,manuscript=letter]{achemso}

\usepackage{chemformula} 
\usepackage{amsmath}
\usepackage{inputenc}
\usepackage{textcomp}
\usepackage{graphicx}
\usepackage{breqn}
\usepackage[T1]{fontenc} 
\usepackage{inputenc}

\author{Adarsh B Vasista}
\email{avasista@ethz.ch}
\affiliation[Unknown University]
{Department of Physics and Astronomy, University of Exeter, Exeter EX44QL, United Kingdom}
\altaffiliation{ Nanophotonics Systems Laboratory, ETH Zurich, 8092 Zurich, Switzerland}

\author{William L Barnes}
\email{w.l.barnes@exeter.ac.uk}
\affiliation{Department of Physics and Astronomy, University of Exeter, Exeter EX44QL, United Kingdom}

\title{Strong Coupling of Multimolecular Species to \textit{Soft Microcavities}}


\begin{document}

\begin{tocentry}
\includegraphics{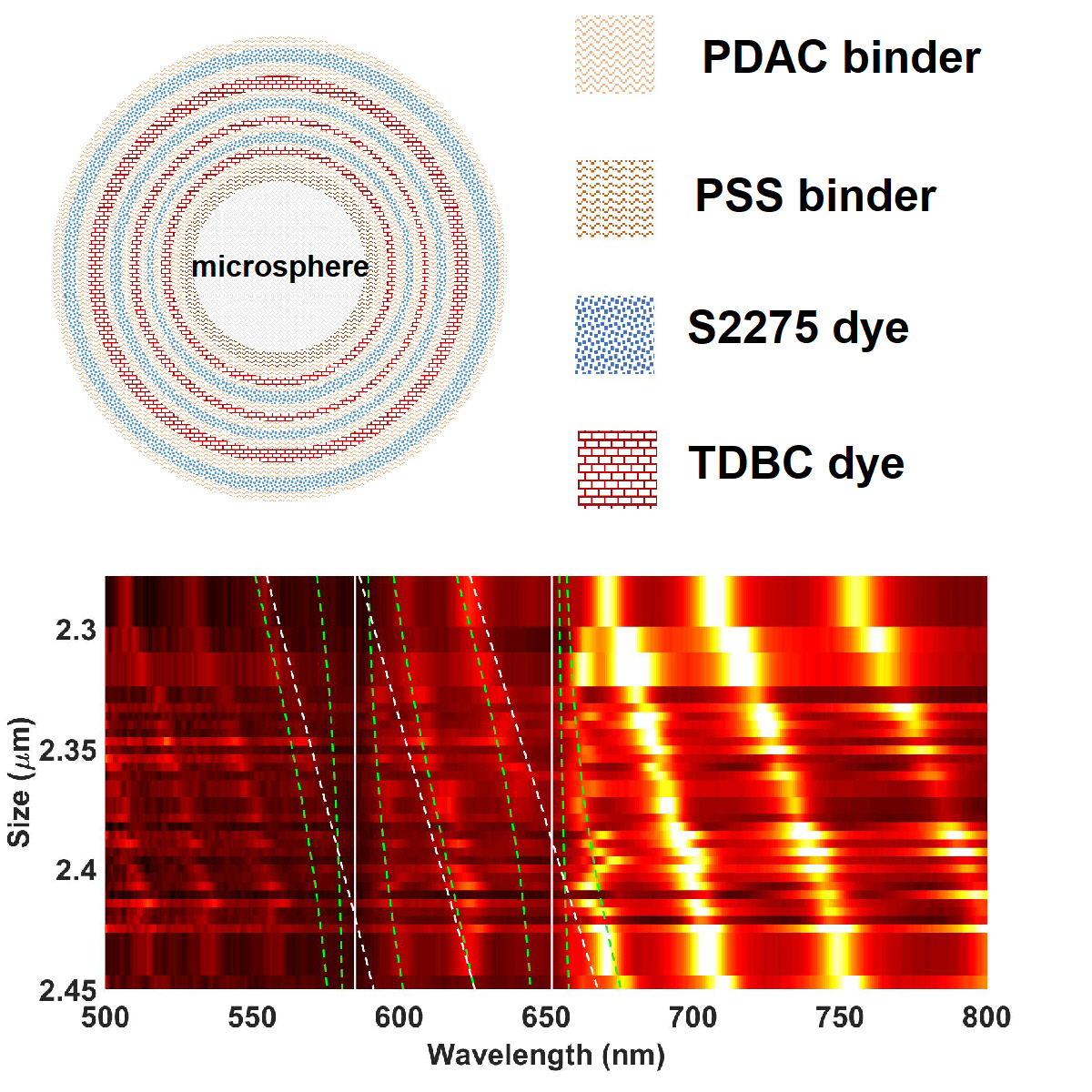}


\end{tocentry}

\begin{abstract}
    Can we couple multiple molecular species to soft-cavities? The answer to this ques-
tion has relevance in designing open cavities for polaritonic chemistry applications. Due
to the differences in adhesiveness it is difficult to couple multiple molecular species to
open cavities in a controlled and precise manner. In this letter, we discuss
the procedure to coat multiple dyes, TDBC and S2275, using a layer-by-layer deposition technique onto a dielectric microsphere so as to facilitate the multi molecule coupling. We observed the formation
of a middle polariton branch due to the inter-molecular mixing facilitated by the whispering gallery modes. The coupling strength,2g, of the TDBC molecules were found
to be 98 meV while that of S2275 molecules was 78 meV. The coupling strength was
found to be greater than the cavity linewidth and the molecular absorption linewidth
showing the system is in the strong coupling regime.

\end{abstract}

Coupling molecules to cavities has  wide implications in controlling molecular properties\cite{4,5}. If the molecule-cavity coupling strength is greater than the losses in the system then the system is said to be in the strong coupling regime, where new energy eigen-states are created\cite{6}. These new states derive their properties from both the cavity and the molecule, thus acting as a platform to engineer optical\cite{7,8}, electronic\cite{9}, and chemical properties\cite{10} of the hybrid system. Various cavity architectures have been studied in the context of strong coupling such as Fabry-Perot cavities\cite{11}, gap plasmon cavities\cite{13}, and single nanostructures\cite{12}. A recent development in the field of molecular strong coupling is the use of dielectric cavities in place of metallic ones\cite{14,15,16}. Dielectric cavities provide sharper cavity resonances and do not suffer from Joule heating losses. Among the different classes of dielectric cavities, microspheres -- also called \textit{soft-cavities} -- have gained prominence recently\cite{2}.

\begin{figure*}
    \centering
    \includegraphics[scale=0.7]{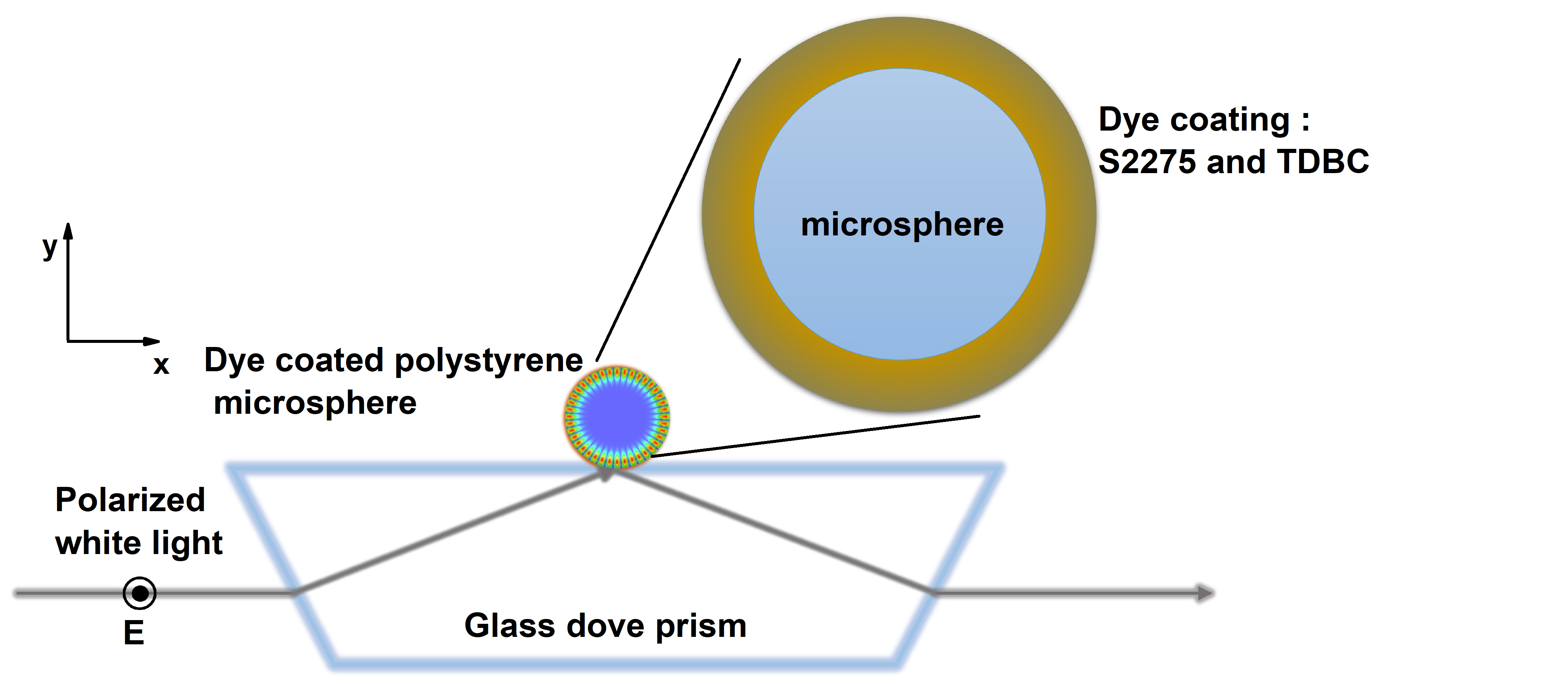}
    \caption{Schematic of the experiment. Individual microspheres were coated with two dye molecules, S2275 and TDBC, consecutively using a layer-by-layer approach. The molecule coated microspheres were then probed optically using evanescent excitation and dark-field microscopy.}
    \label{fig:my_label}
    \end{figure*}

\textit{Soft-cavities} belong to a class of open-cavities and can be easily fabricated, functionalized, are  biocompatible, and can be used in microfluidic environments. They support spectrally sharp resonances called the whispering gallery modes (WGMs). Dielectric \textit{soft-cavities} have been shown to influence molecular emission by changing the polarization and the direction of the emission\cite{17}. They have also been used to detect single molecules\cite{18}, and to strongly couple a molecular monolayer\cite{2}. 

In the past Fabry-Perot cavities\cite{11} and nanoparticles\cite{19} have been utilized to couple multiple molecular species, thereby facilitating inter-molecular coupling. Such inter-molecular coupling is important in harnessing the potential of strong coupling, particularly in areas such as long-range energy transfer\cite{20}, and polariton lasing\cite{21}. To be able to couple multiple molecular species, coupling must take place over an extended spectral range, this can be achieved in two ways. First, a wide spectral range can be achieved if the optical mode employed is also broad, as is the case for example with (non-dispersive) particle plasmon resonances. However the wider linewidth of these particle modes poses significant limitations on the coupling strength. Second, relatively sharp optical modes can be used if they span a wide range because they are dispersive, as is the case for the modes of a Fabry-Perot cavity. However, such modes are ‘closed’ and do not provide dynamic access to the molecular medium. In this context, soft-cavities have an advantage as they are open-cavities and support multiple spectrally sharper resonances.


With this in mind we studied strong coupling of layers of two dye molecules, J-aggregated 5,5',6,6'-tetrachloro-1,1'-diethyl-3,3'-di(4-sulfobutyl)-benzimidazolocarbocyanine (TDBC) and 5-chloro-2-[3-[5-chloro-3-(4-sulfobutyl)-3H-benzothiazol-2-ylidene]-propenyl]-3-(4-sulfobutyl)-benzothiazol-3-ium hydroxide, inner salt, triethylammonium salt (S2275), to individual microspheres of size $\sim$ 2 $\mu$m. 

A schematic of the system under study is shown in figure 1. We consider individual polystyrene microspheres of size $\sim$ 2 $\mu$m coated with a mixture of S 2275 and TDBC dye molecules. The microsphere was placed on a glass doveprism and excited using wideband illumination through the substrate. The evanescent-based illumination excites WGMs inside the microsphere. The scattered light from the sphere in the air medium was captured and analyzed, results will be discussed below.
We first performed finite-element method (FEM) based numerical simulations to understand the molecule-WGM coupling using COMSOL Multiphysics. The wavelength dependent refractive index of the polystyrene microsphere was taken from the literature \cite{1}. The dye coating was modelled as a set of Lorentzian oscillators with permittivity given by,
\begin{equation}
    \epsilon_{dye}(E)=\epsilon_{inf}+\sum_{j=TDBC,S2275}\frac{f_{j}E^2_{j}}{E^2_{j}-E^2-iE\gamma_{j}}
\end{equation}
where $\epsilon_{inf}$ was the background permittivity, set to 1.9, $E_{j}$ was the resonance energy of the dye ($E_{TDBC}$=2.1139 eV and $E_{S2275}$=1.90160 eV), $f_{j}$ was the reduced oscillator strength ($f_{TDBC}$=0.3 and $f_{S2275}$=0.15), and $\gamma_{j}$ was the resonance linewidth of the dye ($\gamma_{TDBC}$=53 meV and $\gamma_{S2275}$=67 meV). The value of $f$ was adjusted so as to match the experimental data. The molecular coating was considered to be homogeneous and 12 nm thick (each layer of the dye was taken as 2 nm thick)\cite{2,22}. 
    \begin{figure*}
    \centering
    \includegraphics[scale=0.8]{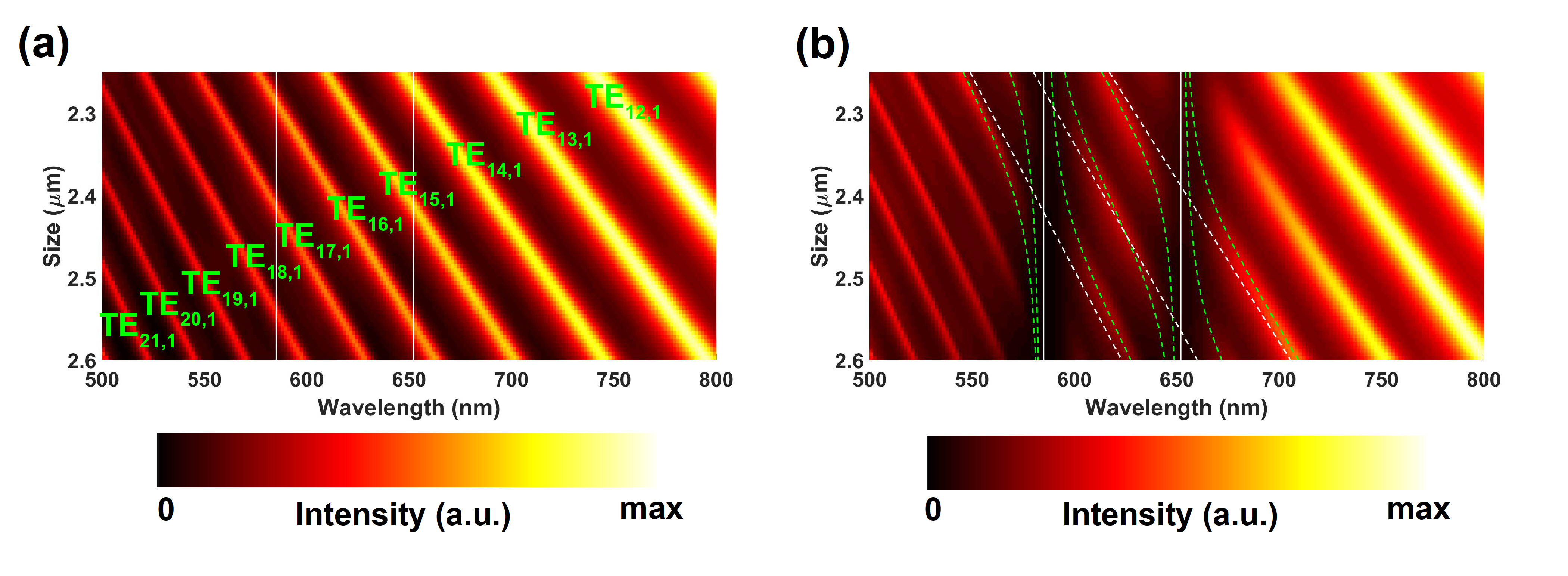}
    \caption{(a) Numerically calculated dispersion of WGMs of \textit{bare} microspheres of size $\sim$ 2 $\mu$m placed on a glass substrate. (b) Numerically calculated dispersion of WGMs of microspheres of size $\sim$ 2 $\mu$m coated with 3 layers of S2275 and TDBC molecules. The superimposed dashed green lines represent the eigen values of the Hamiltonian calculated to fit the experimental and numerical data. The dashed white line represents the uncoupled WGM and the solid white lines represent molecular resonances. }
    \label{fig:my_label}
\end{figure*}

Figure 2 (a) shows the numerically calculated dispersion of the scattering spectra of uncoated individual microspheres, the spectrally sharp WGMs of the microsphere are clearly evident. Each WGM is characterized by two mode numbers $(m,n)$. The azimuthal mode number $m$ is half the number of electric field maxima along the periphery of the sphere while the radial mode number $n$ is the total number of electric field maxima along the radius of the sphere. (Please refer to \cite{2} for more details on assigning the mode numbers.) The polarization of light was kept as transverse electric (TE). Figure 2 (b) shows the dispersion of individual microspheres coupled to three layers of PDAC/TDBC and S2275 dye. We can clearly see that the modes TE$_{17,1}$, TE$_{16,1}$, and TE$_{15,1}$ were affected by the presence of the molecular absorption. New energy eigen-states called polaritons were formed by mixing multiple WGMs and molecular resonances through the molecule-WGM coupling. 

The Hamiltonian of the total system can be described as\cite{3},
\begin{equation}
    \mathcal{H}=\bigoplus\limits_{j=1}^{3}
    \begin{pmatrix}
    E_{PDAC/TDBC}-j\frac{\gamma_{PDAC/TDBC}}{2} & 0 & -g_{j,TDBC}\\
    0 & E_{S2275}-j\frac{\gamma_{S2275}}{2} & -g_{j,S2275}\\
    -g_{j,TDBC} & -g_{j,S2275} & E_{j}-j\frac{\gamma_{j}}{2} 
    \end{pmatrix}
\end{equation}

\noindent where $E_{j}$ represents the energy of the $j^{th}$ WGM of the resonator and $\gamma_{j}$ represents its linewidth. The Eigenvalues of equation 2 gives us the polariton energies while the eigen vectors provide an estimate of the mixing fractions (Hopfield coefficients). We fit the numerical simulation data with the eigenvalues from equation 2 and the results are superimposed on figure 2 (b). The value of the coupling strengths were found to be $g_{TE_{17,1}, PDAC/TDBC}$ = 46 meV, $g_{TE_{17,1}, S2275}$ = 0 meV, $g_{TE_{16,1}, PDAC/TDBC}$ = 49 meV, $g_{TE_{16,1}, S2275}$ = 39 meV, $g_{TE_{15,1}, PDAC/TDBC}$ = 0 meV, $g_{TE_{15,1}, S2275}$ = 39 meV. In the cases where WGMs interact with the molecular resonance ($TE_{17,1}$ with PDAC/TDBC, $TE_{15,1}$ with S2275, and $TE_{16,1}$ with PDAC/TDBC and S2275) the coupling strength, $2g$, was found to be greater than the mean of the coupling molecular resonance linewidth ($\gamma_{TDBC}$=53 meV, $\gamma_{S2275}$=67  meV ) and the WGM linewidth ($\gamma_{TE_{17,1}}$=13 meV, $\gamma_{TE_{16,1}}$=15 meV, $\gamma_{TE_{15,1}}$=21 meV), so that the strong coupling regime has been reached \cite{23}. This shows that \textit{soft-cavities} support spectrally sharp resonances which can be utilized to strongly couple multiple molecular species. 

\begin{figure*} 
    \centering
    \includegraphics[width=\linewidth]{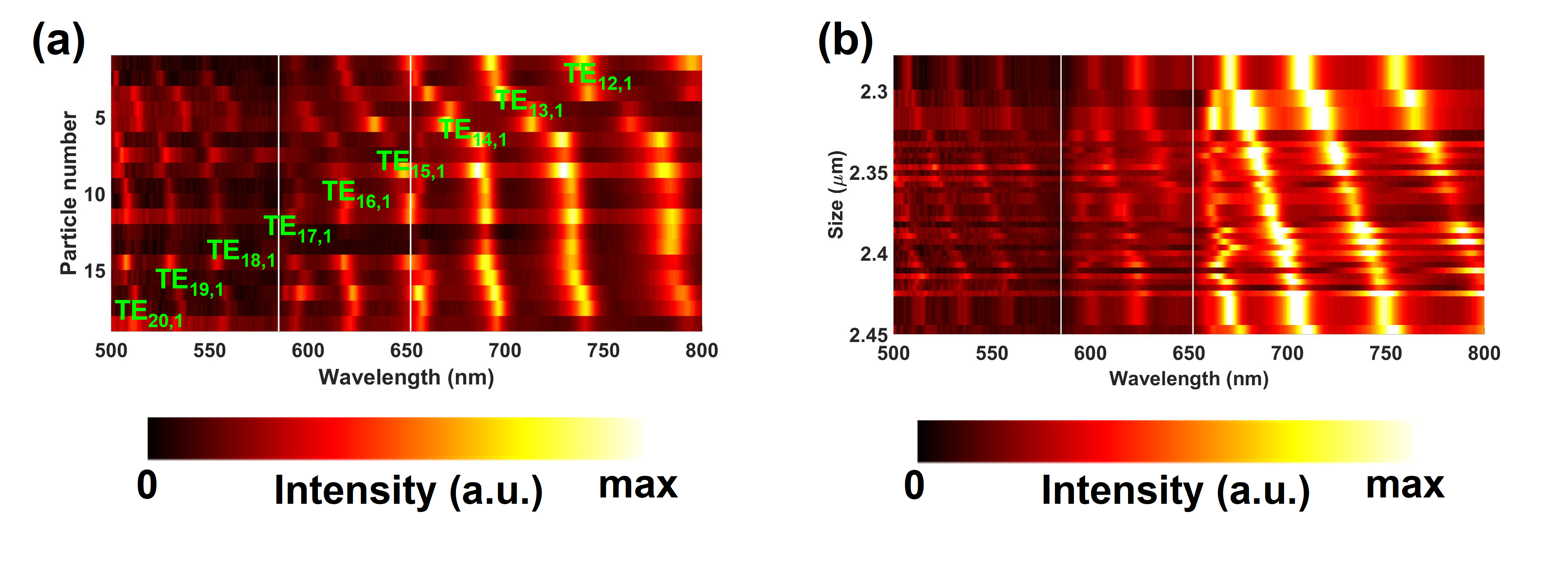}
    \caption{(a) Experimentally measured dispersion of the microspheres coated by the mixture of S2275 and TDBC dye molecules. The size of the microspheres were around 2 $\mu$m  (b) Experimentally measured dispersion of the microspheres coated with alternative layers of S2275 and TDBC molecules. The solid white lines represent the positions of molecular resonances. }
    \label{fig:my_label}
\end{figure*}
Having established strong coupling of multi-molecular species to \textit{soft-cavities} numerically, we now focus on the experimental implementation using a layer-by-layer deposition (LBL) method. In a typical deposition step we mixed 20 $\mu$L of anionic polystyrene sulfonate (PSS) solution (20$\%$ by weight in water, diluted 1:1000) with 1 mL of polystyrene microsphere colloidal solution (15\% by weight in water, diluted 1:50) and the resulting solution was allowed settle for 20 min. The solution was then washed three times in water to remove excess polyelectrolyte solution. This step was followed by mixing  cationic polyelectrolyte Poly-(diallyl-dimethylammoniumchloride) (PDAC) (20$\%$ by weight in water, diluted 1:1000), the resulting solution was also allowed to settle for 20 min. The solution was washed three times in water to remove excess polyelectrolyte. The dye mixture was prepared by mixing 0.5 mL of 0.01 M TDBC solution and 0.5 mL of 0.01 M S2275 solution. Then, 40 $\mu$L of this dye mixture was mixed with the polyelectrolyte coated microsphere solution and allowed to settle for 20 min. The solution was washed with water three times to remove excess dye solution. This step was repeated with PDAC as binder layer a further five times. 

The individual dye coated microspheres were then probed using evanescent excitation in a dark-field configuration. White light was used to excite WGMs of the microsphere and the scattered light was then collected using a 0.8 NA, 100x objective lens. To obtain a dispersion plot, we collected spectral signatures from multiple microspheres, each with a slightly different size, and then ordered the spectra in ascending order of the size of the microspheres (See section S1 of supplementary information for details of experimental setup).  We excited the spheres with TE polarized light as the deposited molecular layers show zero radial dipole moment and hence the TM polarized scattering shows no signs of molecule-cavity coupling. \cite{2}

Figure 3 (a) shows the experimentally measured dispersion of the microspheres after coating the mixture of dyes . The dispersion was created by arranging individual dark field scattering spectra from 20 microspheres in ascending order. We can see that the WGM in resonance with TDBC absorption (TE$_{17,1}$) was perturbed due to TDBC-cavity coupling. However the mode resonant with S2275 absorption (TE$_{15,1}$) shows almost no change. This clearly shows that mixing two dyes to form a solution and then coating the mixture on a microsphere does not yield multi molecule - cavity coupling. We suggest that this happens because of the different polarity of the dyes. LBL is a charge based deposition technique where the anionic dye is deposited as a layer due to the electrostatic attraction between polyelectrolyte binder and the dye. The total charge carried by TDBC dominates the dye mixture and hence TDBC is selectively adsorbed on the microsphere and hence we see a molecule-cavity coupling signature only from TDBC and not from S2275. Note that it is difficult to calculate the exact size of the microspheres, in this case, by fitting the measured spectra. This is due to the unknown percentage of the S2275 and TDBC molecules adsorbed on the microspheres; that is why the dispersion is shown as a function of particle number. However the overall message of molecule-cavity coupling can be conveyed effectively using particle number rather than size.  


To solve the problem of adhesiveness, we modified the deposition procedure by alternatively depositing TDBC and S2275 molecules on to the microsphere. The polyelectrolyte deposited microsphere solution was mixed with 20 $\mu$L of 0.01 M TDBC dye solution and allowed to settle for 20 minutes. The solution was then washed in water three times to remove excess TDBC molecules. Then a layer of PDAC as a binding layer was deposited using the procedure described earlier. Then we mixed 20 $\mu$L of 0.01 M S2275 dye solution and allowed it to settle for 20 minutes. The solution was then washed in water three times to remove excess dye molecules. This procedure was repeated to deposit three layers of TDBC and S2275 each. Dye coated microspheres were then dropcast on a glass substrate and probed with evanescent excitation dark-field spectroscopy.

Figure 3 (b) shows the experimentally measured dispersion of the WGMs of microspheres after coating them with alternative layers of dye molecules. Now we can see, in contrast to figure 3 (a), splitting and anti-crossing of modes TE$_{15,1}$ and TE$_{17,1}$ which are resonant with S2275 and TDBC absorption respectively. The size of microspheres in this case was calculated by fitting the measured scattering spectra with the numerically calculated spectra. To further analyze the coupling of the two types of dye molecules to an individual microsphere, we fit the experimental data with a simple coupled oscillator model, given by equation 2, as shown in figure 4 (a).

\begin{figure*} 
    \centering
    \includegraphics[scale=0.8]{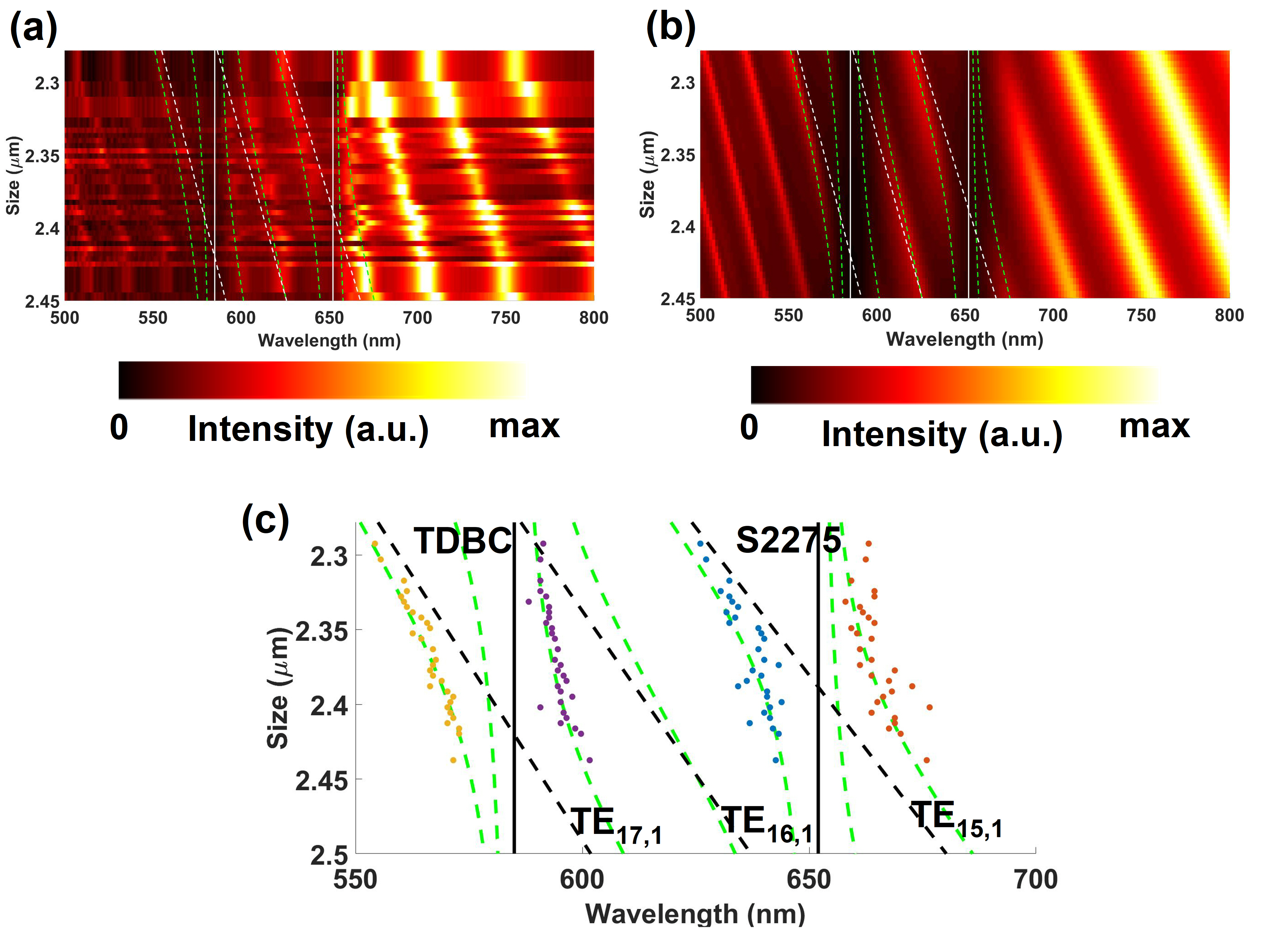}
    \caption{(a) Experimentally measured dispersion of WGMs of microspheres of size $\sim$ 2 $\mu$m coated with 3 layers of S2275 and TDBC molecules. (b) Numerically calculated dispersion of WGMs of microspheres of size $\sim$ 2 $\mu$m coated with 3 layers of S2275 and TDBC molecules. The superimposed dashed green lines represent the eigen values of the Hamiltonian calculated to fit the experimental and numerical data. The dashed white line represents the uncoupled WGM and the solid white lines represent molecular resonances. (c) Calculated dispersion plot of the mode TE$_{15,1}$, TE$_{16,1}$, and TE$_{17,1}$ using a coupled oscillator model to fit to the experimental data. The experimental values of the spectral positions of the lower and upper polaritons were extracted from Figure 4a and are represented as dots. The dashed green lines are the fit using a coupled oscillator model. }
    \label{fig:my_label}
\end{figure*}

The size range of the available microspheres was limited (see figure 2 and figure 4) due to manufacturing limitations. Nonetheless we can see clear splitting and anticrossing of the modes TE$_{17,1}$, TE$_{15,1}$ which are spectrally resonant with the molecular absorption of the PDAC/TDBC and S2275 dye molecules respectively. Figure 4 (b) is the numerically calculated dispersion for the size range of 2.25 $\mu$m - 2.45 $\mu$m nicely matching the experimental data. We superimpose the calculated eigenvalues of the Hamiltonian on the experimental data of figure 4 (a). We used the same parameters for the Hamiltonian to fit the experimental as well as the numerical data. A discussion on the mixing fraction, also called Hopfield coefficients, is given in section S2 of the supplementary information.The width of the LPB and UPB created due to strong coupling of TDBC molecules to TE$_{17,1}$ was found to be 23$\pm$1 meV and 34$\pm$1 meV respectively. In the case of polaritons formed due to strong coupling of TE$_{15,1}$ to S2275, the width of LPB was found to be 30$\pm$1 meV and that of UPB was found to be 35$\pm$1 meV. We calculated the widths of polariton branches by fitting the scattering spectra of the microsphere of size 2.34$\mu$m. To make the strong coupling of multi molecular species to WGMs clearer we extract the spectral positions of the LPB and the UPB from figure 4 (a) and plot them as a function of the size of the microspheres in figure 4 (c). We also fit the experimental data using the coupled oscillator hamiltonian defined in equation 2. Figure 4 (c) provides clarity in noticing the splitting and anti-crossing between the polariton branches.   

Similar results have been obtained with the Fabry-Perot resonators\cite{11}, and with nanoparticles\cite{19}. An important difference between Fabry-Perot resonators and \textit{soft-cavities} is that the former shows well defined angular dispersion of the modes while the latter has no angular dispersion. In the case of $soft-cavities$, the dispersion of the modes is a \textit{collective effect} defined by the size variation of the resonator (as the spectral positions of the WGMs are size dependent). Since we chose to work with a single microsphere, in any given implementation we will only be probing a limited range of the total range of mixing fractions, unlike the case for the Fabry-Perot resonator. This indicates that the equivalence of size in the case of \textit{soft-cavities} with the angle of incidence in the Fabry-Perot resonators is not a complete equivalence. 

To summarize, we have numerically and experimentally demonstrated strong coupling of multiple molecular species with dielectric \textit{soft-cavities} through dark-field scattering signatures. We used layer by layer deposition method to accurately control the molecular deposition on the microspheres. We also discussed the procedure to deposit multiple dyes onto the microspheres to facilitate multi molecule strong coupling. We anticipate that these results will find relevance in designing open-cavities for polariton mediated multi molecular interactions such as energy transfer, lasing etc. As the microspheres can be trapped and moved in a microfluidic environment, the results discussed here can be extrapolated to achieve multi-molecule strong coupling in a dynamic microfluidic environment.  

\section*{Supporting Information}
Additional experimental details including a schematic of the setup and a discussion on the Hopfield coefficients. 
\section*{Acknowledgements}
The authors acknowledge the support of European Research Council through the Photmat project (ERC-2016-AdG-742222: http://www.photmat.eu). ABV thanks Wai Jue Tan and Kishan Menghrajani for their help in the preparation of the samples. Research data are available from the University of Exeter repository at https://doi.org/...
\bibliography{ref_jpc}
\bibliographystyle{}
\end{document}